\documentclass[marginparwidth=30pt,a4paper,12pt]{article}

\usepackage[utf8]{inputenc}
\usepackage{graphicx}
\usepackage{amsmath}
\usepackage{gensymb}
\usepackage{multirow}
\usepackage{cleveref}
\usepackage{caption}
\usepackage{float}
\usepackage{xcolor}
\usepackage{authblk}
\usepackage{appendix}

\usepackage[style=nature]{biblatex} %Imports biblatex package
\usepackage{comment}

\title{Failure of topologically interlocked structures - a Level-Set-DEM approach}
\author[1]{Shai Feldfogel}
\author[2]{Konstantinos Karapiperis}
\author[3]{Jose Andrade}
\author[1]{David S. Kammer}
\affil[1]{Institute for Building Materials, ETH Zurich, Switzerland}
\affil[2]{Department of Mechanical and Process Engineering, ETH Zurich, Switzerland}
\affil[3]{Department of Mechanical and Civil Engineering, Caltech, Pasadena, California, USA}

\begin{document}

\begin{comment}

Why (purpose)
Convince people that our approach to modeling TIS is principally superior to the hitherto used FEM, and that therefore it, not FEM, should be the basis for new and improved models.

Who (audience): anybody doing TIS, TIM, Discrete structure, structural andlysis with DEM, LS-DEM, DEM

What (Main messages)

0 (and most important) All of you (audience): you should use our model!

1) Unlike standard FEM-based approaches, our new computational model captures the entire failure process in TIS - A necessary condition to understand, and eventually be able to predict this complex phenomena.
2) Our model also provides better quantitative estimates of the structural response (load deflection curve, global stiffness, carrying capacity, toughness) and compared with standard tools.
3) Our model can follow the evolution of contacts through the structural response, up to failure, and thus shed light on the local interfacial mechanics that govern the behavior and failure of TIS.
4) Therefore, it makes more sense to address of the failure of TIS using our approach over FEM.

\end{comment}

\maketitle

\begin{abstract}
Topological Interlocking Structures (TIS) are assemblies of interlocking building blocks that hold together solely through contact and friction at the blocks' interfaces, and thus do not require any connective elements. 
This salient feature makes them highly energy-absorbing, resistant to crack propagation, geometrically versatile, and re-usable. 
It also gives rise to failure mechanisms that, differently from ordinary structures, are governed by multiple contact interactions between blocks and frictional slip at their interfaces.
Commonly-used modeling tools for structural analysis severely struggle to capture and quantify these unusual failure mechanisms.
Here, we propose a different approach that is well suited to model the complex failure of TIS. 
It is based on the Level-Set-Discrete-Element-Method, originally developed for granular mechanics applications.
After introducing the basic assumptions and theoretical concepts underlying our model, we show that it well-captures experimentally observed slip-governed failure in TIS slabs and that it estimates the force-displacement curves better than presently available modeling tools.
The theoretical foundation together with the results of this study provide a proof-of-concept for our new approach and point to its potential to improve our ability to model and to understand the behavior of interlocked structural forms.
\end{abstract}

\section{Introduction} \label{sec:Introduction}

\begin{figure} 
    \includegraphics[width=1\textwidth]{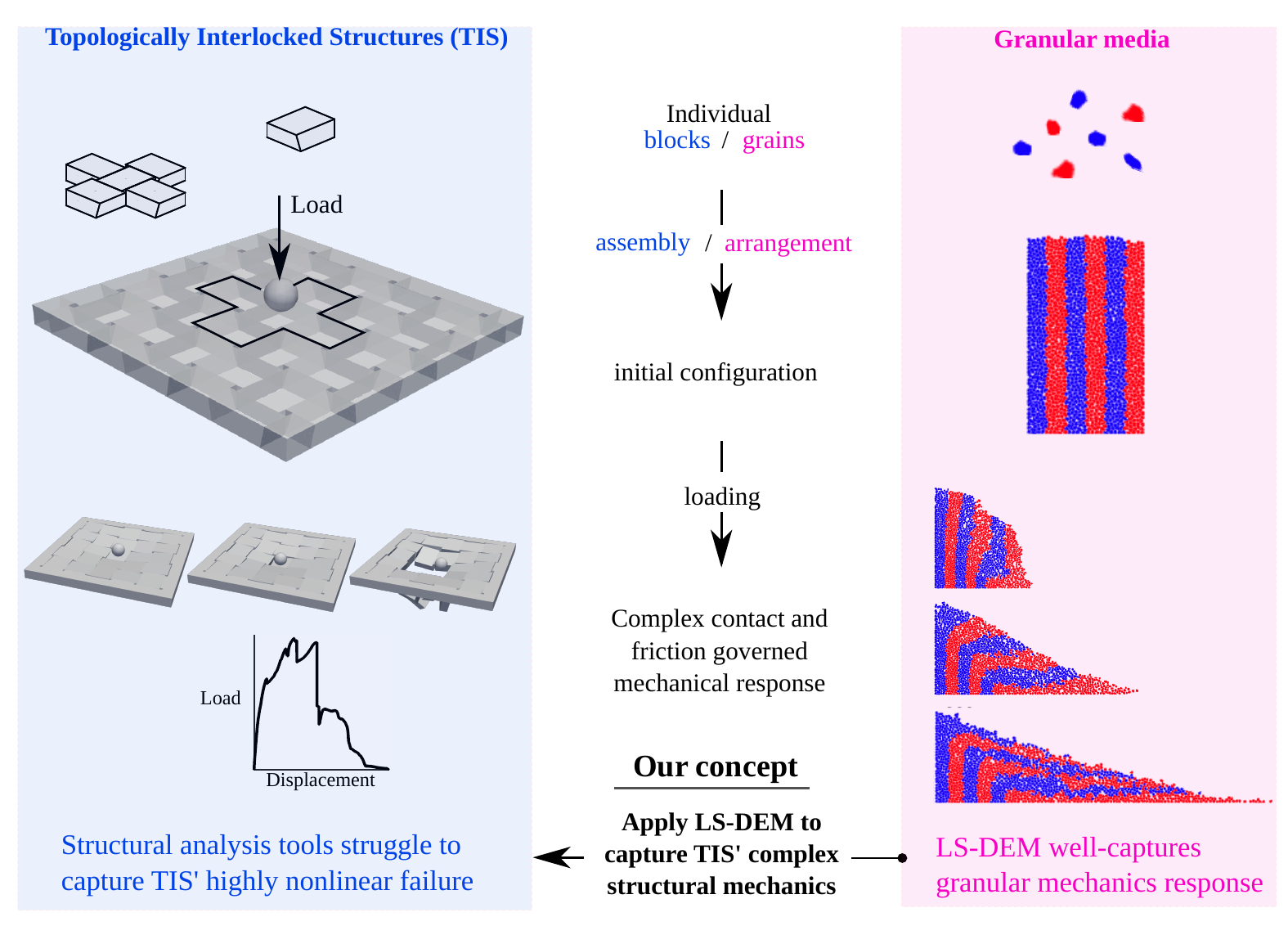} 
    \centering
    \caption{Illustration of presented concept: Based on the similarities between TIS and granular media on the one hand, and LS-DEM unique ability to model the latter's mechanics on the other, we apply LS-DEM to model the complex failure of TIS, which common structural analysis tools struggle to capture.}
    \label{fig:Introduction_figure}
\end{figure}

% What are TI structures, what is special about them, and why they are useful
Topological Interlocking Structures (TIS) are assemblies of interlocking building blocks that hold together solely through contact and friction at the blocks' interfaces, and thus do not require any connective elements, see Fig. \ref{fig:Introduction_figure} left. 
This defining feature sets them apart from ordinary structural forms and it is responsible for their unique behavior and advantageous properties. 
These include high energy absorption, high resistance to crack propagation, large tolerance to missing blocks and to geometrical imperfections, geometrical versatility, re-usability, and more \cite{dyskin_principle_2005,molotnikov_percolation_2007, carlesso_enhancement_2012, carlesso_improvement_2013,dyskin_topological_2019,dyskin_mortarless_2012}. 
% TI structures have yet to become widespread in spite of their potential because we don't understand well enough how they behave and fail
In spite of these useful properties, TIS's promising potential is yet to translate to large-scale prevalence. 
One likely reason is that our ability to predict their failure is far from fully developed. 
As a result, the ability to design them safely, a prerequisite for widespread application, is limited.

% The failure of TI structures is challenging to understand and quantify because of their unique structural behavior, which is essentially different from the ordinary structures. TI behave midway between ordinary structures and granular media
Developing predictive capabilities for the behavior and failure of TIS is challenging because the blocks of TIS are, in general, not connected by any mechanical means (e.g., adhesives or bolts). 
This means that the structural integrity depends entirely on the transmission of forces through the interfaces, and, these interfacial forces are difficult to quantify and predict because the interfacial contact conditions that govern them are: (a) geometrically irregular and dynamically changing by nature \cite{djumas_deformation_2017}; (b) highly dependent on local slip failures \cite{djumas_deformation_2017,mirkhalaf_toughness_2019,koureas_failure_2022}; (c) coupled with all other interfaces through the global response; and (d) sensitive to unavoidable geometrical imperfections \cite{mirkhalaf_toughness_2019,barthelat_novel_2011}.

% The main questions that arise due to the unique load transfer mechanisms, and that require addressing to better understanding are:
As shown and discussed ahead, commonly-used models struggle to capture and quantify the slip-governed failure of TIS, underscoring a need for alternative approaches.
The main objective of this study is to establish a proof-of-concept for a new computational approach that is better equipped to model the failure of TIS than presently available tools.
The basic idea underlying our approach is to apply the Level-Set-Discrete-Element-Method (LS-DEM), originally developed for granular applications, to structural analysis of TIS, see Fig. \ref{fig:Introduction_figure}.

The most commonly-used tool to model the behavior and failure of TIS is the Finite Element Method (FEM), see \cite{williams_mechanics_2021,short_scaling_2019,djumas_deformation_2017,mirkhalaf_toughness_2019,schaare_point_2008, dalaq_manipulating_2020, dalaq_strength_2019}. 
In cases where the response was entirely governed by a stick regime and the specimens were not loaded up to failure, FEM obtained a very good agreement with experimental and analytical results\cite{schaare_point_2008,short_scaling_2019}. 
In the context of beam-like assemblies \textit{with few blocks}, FEM was also able to correctly capture the experimentally observed slip-governed failure mechanism and match well the global load displacement curves \cite{dalaq_manipulating_2020}. 
However, as stated in \cite{dalaq_manipulating_2020}: "we have limited the study to 5 blocks (N = 5) because of expensive computational costs with larger N".
FEM's difficulty with handling more than a few blocks becomes a major obstacle in the context of the most common TIS application - slabs, which typically comprise dozens of blocks.
FEM's difficulty with TIS slabs is expressed, for example, by over-prediction of the peak load by an order of magnitude in \cite{mirkhalaf_toughness_2019}, and by divergence of the analyses from the experimental results close to failure and an inability to capture the experimentally observed load drops \cite{djumas_deformation_2017}. 
In general, FEM struggles to capture the experimentally observed slip-governed failure in TIS slabs and to follow the corresponding load-displacement response up to failure.
Since these capabilities are important to properly model TIS failure, alternatives to FEM are warranted.

The Discrete Element Method (DEM) was originally designed to model dynamically-evolving contact and friction interactions between multiple spherical grains \cite{cundall_discrete_1979}. 
As such, it is a natural starting-point framework for a model that could better address the intricate behavior and failure of TIS better.
The ability of DEM to handle multiple dynamic contacts is due to an explicit dynamic framework, a rigid-body assumption, elementary block shapes (mostly spherical, and generally convex), and a penalty-enforced contact between the blocks. 
The potential of a DEM-based approach for TIS is supported by the 3D FEM analysis in \cite{schaare_point_2008}.
There, excellent agreement with experimental results was obtained using extremely coarse meshes of only 8 elements per block (three orders of magnitude less than in \cite{djumas_deformation_2017}).
This suggests that a coarse representation of block deformation, one that is possible even under the seemingly contradictory rigid body assumption as will be explained ahead, may be sufficient to capture the essential features in the behavior of TIS.
However, differently from a coarse meshed FEM approach, which would lack the spatial resolution of the contact kinematics necessary to capture stick-slip transition and the slip regime, these pose no special difficulties for a DEM-based model.
DEM was used by Brugger et al. \cite{brugger_experiments_2008,brugger_numerical_2009} to model centrally loaded TIS slabs with cube shaped blocks, but this approach has not been further explored.
The essential limitation of ordinary DEM as a general modeling approach to TIS is the inability to fully address the geometrical variety of TI blocks and their complex contacts. 
Recent DEM variants were developed that can handle non-convex polygonal blocks in the context of granular flow \cite{govender_hopper_2018}.
However, these variants lack the ability to represent the geometry of curved faced blocks, such as the popular osteomorphic blocks, \cite{djumas_deformation_2017,djumas_enhanced_2016,dyskin_principle_2005,estrin_design_2021,estrin_architecturing_2021}, and to resolve the conforming contact interactions between such blocks.
As such, they are not fully equipped to address the full range of TIS, and therefore cannot provide a general modeling solution.

Recently, a DEM variant called Level-Set-DEM (LS-DEM) \cite{kawamoto_level_2016} was developed specifically to overcome the shape limitations of ordinary DEM. 
LS-DEM is able to represent arbitrary block geometries and resolve the complex contact kinematics that arise between them through a node-based discretization of block boundary.
This shape versatility, together with the ability to handle non-convex shapes and the aforementioned DEM advantages, makes LS-DEM a potentially attractive approach for assemblies where interlocking comes into play, see Fig \ref{fig:Introduction_figure} right and \cite{karapiperis_stress_2022}. 
Recently, LS-DEM's original contact formulation has been adapted in a way that enabled us to use it for structural analysis \cite{feldfogel_discretization-convergent_2022}.
However, LS-DEM ability to realistically capture and predict the behavior and failure of TIS as observed in experiments - a necessary validation test for a model - has not yet been established.

Summarizing, efficient and reliable computational tools are indispensable to modeling the complex slip-governed failure of TIS.
Yet, the two computational approaches hitherto employed in the literature are insufficient, FEM struggling to capture and quantify the slip-governed failure of TIS, and DEM lacking the necessary ability to handle arbitrarily shaped blocks. 
The shape-versatile LS-DEM can capture the slip-governed failure in TIS with arbitrarily shaped blocks, but it has not yet been validated against experimental results in this context. 
All of this underscores a lack of reliable computational tools which undermines our ability to model TIS and, ultimately, to design them.
The main objective of this manuscript is therefore to establish a proof-of-concept for LS-DEM as a new and improved structural analysis computational tool for TIS.
This proof-of-concept will be established in what follows on theoretical grounds and through validation with available experimental data.

Next, in section \ref{sec:Methodology} we present the methodology, focusing on the physical modeling assumptions, the mathematical formulation, the basic concepts, and the limitations of our LS-DEM model.
In section \ref{sec:Results and discussion}, we apply LS-DEM to analyse the centrally-loaded TIS slabs studied in \cite{mirkhalaf_toughness_2019}, demonstrating its ability to correctly capture and to predict the experimentally observed failure mechanisms better than presently available tools.

\section{Methodology} \label{sec:Methodology}
The assumptions underlying LS-DEM are outlined in section \ref{subsec:Assumptions}; the mathematical formulation is described in section \ref{subsec:Mathematical formulation}; the concept of accounting for deformability under the rigid body assumption is illustrated and explained in section \ref{subsec:Modeling deformability with rigid blocks}; the limitations of the model are discussed in section \ref{subsec:Limitations}.

\subsection{Assumptions} \label{subsec:Assumptions}
The modeling assumptions involve global considerations, the blocks, and the interfaces. 
Globally, the structural response is defined by the 3D rigid body motions of the blocks, which are governed by Newton's generalized laws of motion.
Accordingly, the total number of degrees of freedom equals the number of blocks times six (the number of rigid body degrees of freedom).
The energy dissipation mechanisms comprise sliding friction, restitution losses, and global damping.

The blocks are assumed to be unbreakable rigid bodies; their mass corresponds to their true material density (no mass scaling); and the forces acting on them comprise gravity, contact and friction interface forces by adjacent blocks, support reactions by Dirichlet boundaries, and damping forces. 
The interfaces are assumed to be adhesion-less, so only normal compressive forces and tangential friction forces are considered; contact is enforced in a linear-penalty sense; and a bi-linear Coulomb's friction law where $\mu_{static}=\mu_{kinetic}$ is assumed to govern the tangential forces. 

\subsection{Mathematical formulation} \label{subsec:Mathematical formulation}
The mathematical formulation of LS-DEM has been detailed elsewhere \cite{kawamoto_level_2016,feldfogel_discretization-convergent_2022} and it is not repeated here in full for the sake of brevity.
Nevertheless, the adapted contact formulation introduced in \cite{feldfogel_discretization-convergent_2022} and adopted here is briefly described for completeness.

\begin{figure}[H]
    \includegraphics[width=1\textwidth]{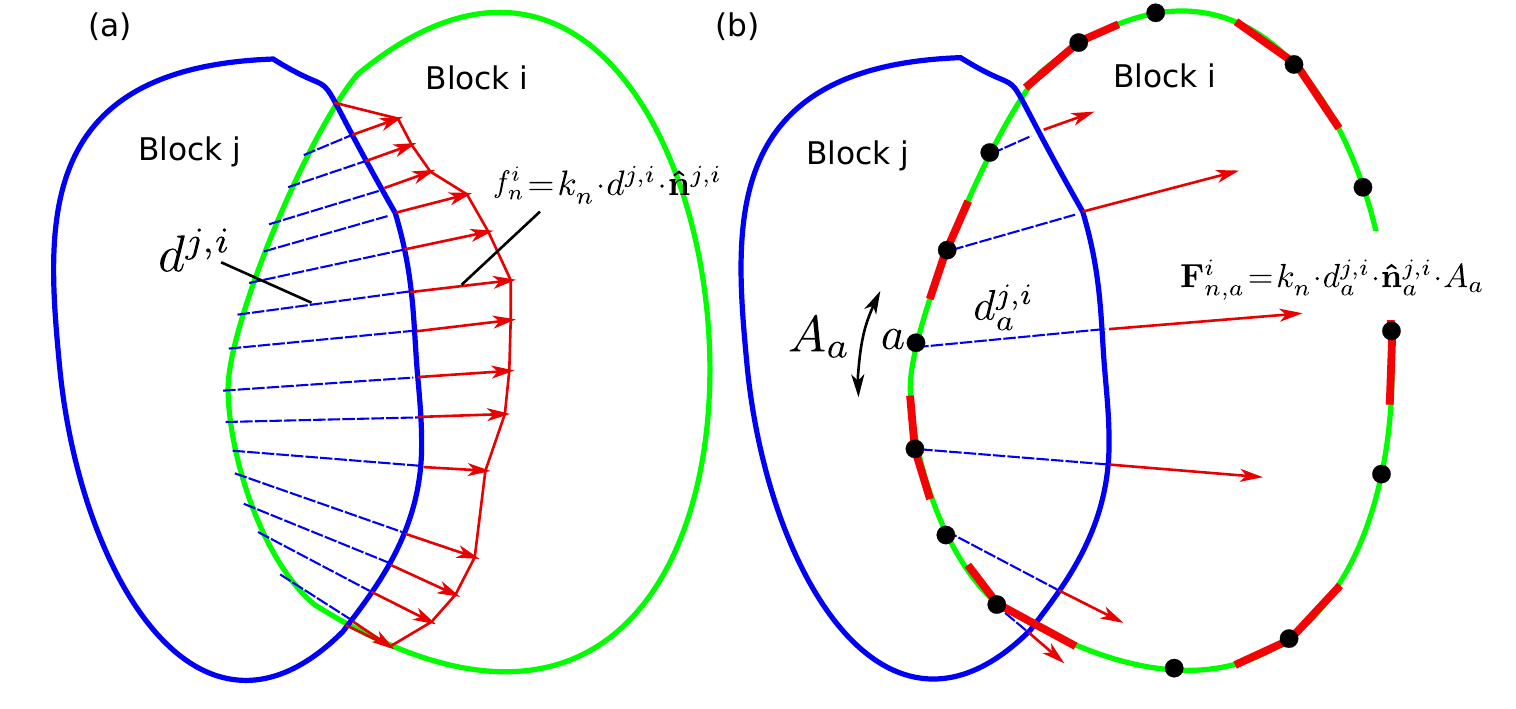} 
    \centering
    \caption{Contact modeling - (a) the continuum-based approach; and (b) LS-DEM's discretized nodal forces (penetrations are grossly exaggerated for illustrative purposes)}
    \label{fig:contact}
\end{figure}

As detailed in \cite{feldfogel_discretization-convergent_2022} and as illustrated in Fig. \ref{fig:contact}(a), we adopt a continuum-based contact approach wherein contacting block surfaces are thought of as elastic foundations, exerting equal and opposite normal compressive tractions $f^{i}_n$ proportional to the penetrations $d^{j,i}$ at each contact point.  
Accordingly, the normal penalty parameter $k_n$ has dimensions of traction per unit penetration and it is analogous to the elastic foundation modulus.

In LS-DEM, the block surfaces are discretized by seeding nodes on them, as schematically shown on block i in Fig. \ref{fig:contact}(b). 
Accordingly, the continuous contact tractions in Fig. \ref{fig:contact}(a) are represented by discrete nodal forces, shown as red arrows in Fig. \ref{fig:contact}(b). 
The nodal force $\mathbf{F}^i_{n,a}$ at contact node $a$ reads\footnote{To avoid redundant symbols, $k_n^*$ from \cite{feldfogel_discretization-convergent_2022} has been denoted here by $k_n$, with the understanding that its dimension is still traction per unit displacement and not force per unit displacement as in the original formulation.}:
\begin{equation} \label{eq:adapted LS-DEM Fna}
    \mathbf{F}^i_{n,a} = k_n \cdot d^{j,i}_a \cdot \mathbf{\hat{n}}^{j,i}_a \cdot A_a
\end{equation}
where the subscript $a$ represent the $a$'th contact node and where $A_a$ is the nodal tributary area.

\subsection{Modeling deformability with rigid blocks} \label{subsec:Modeling deformability with rigid blocks}
Under the rigid body assumption, our model cannot directly account for block deformability, which governs the stiffness and capacity of TIS, through the in-plane stiffness.
Instead, we account for it indirectly through the block penetrations, which are inherent in the penalty contact formulation, as discussed above.
Next, we show how the penalty parameter $k_n$ can be tuned to approximate the actual deformability of the blocks in two cases - the geometrically perfect case and the geometrically imperfect case. 

\noindent{\underline{\textbf{$k_n$ in the geometrically perfect case}}}

\begin{figure}[H] 
    \includegraphics[width=1\textwidth]{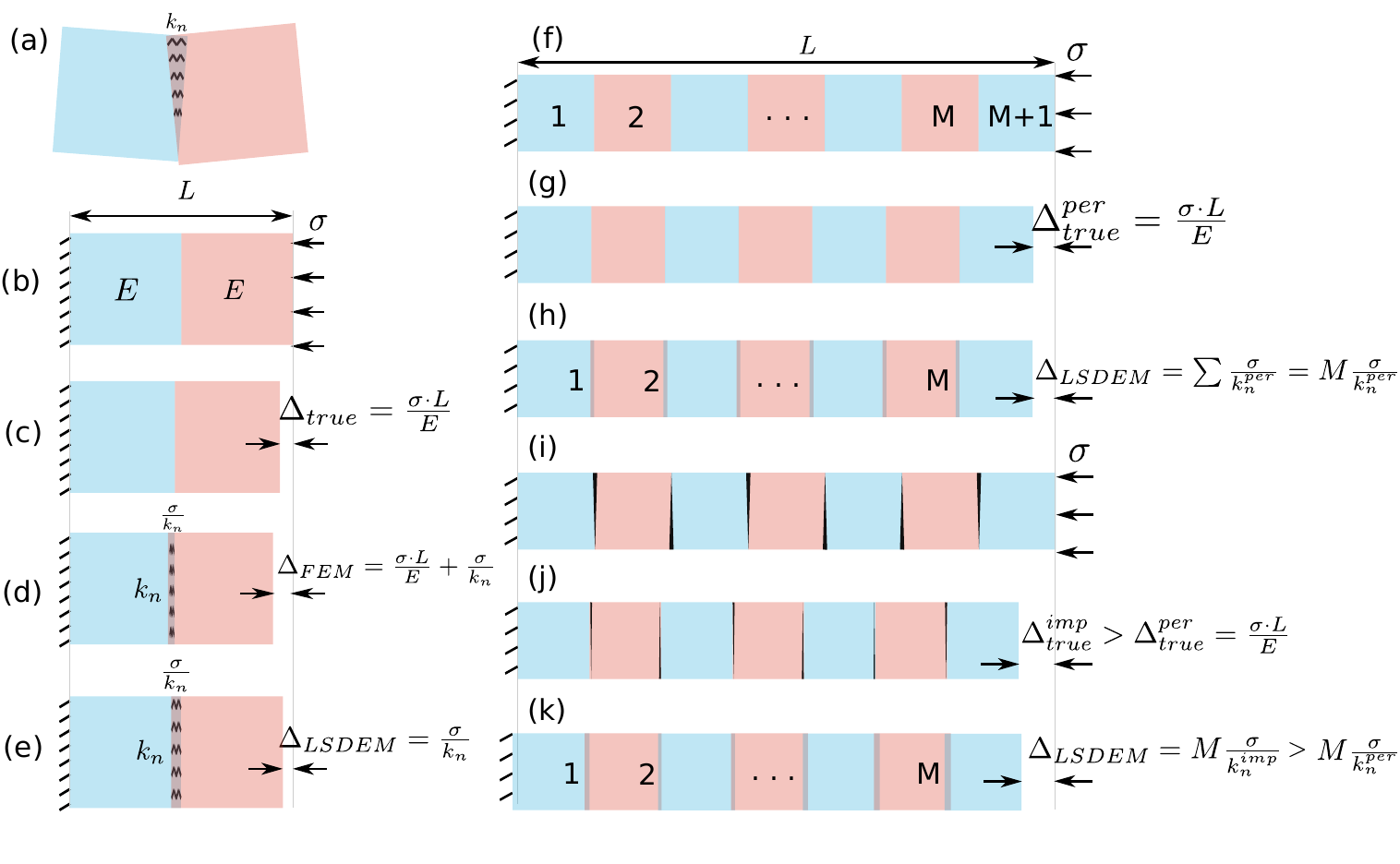} 
    \centering
    \caption{Methodology - the in-plane deformability is accounted for in LS-DEM through interfacial penetrations: (a) penetrating surfaces as elastic foundations with modulus $k_n$; (b-h) the geometrically perfect case; (i-k) the geometrically imperfect case}
    \label{fig:modeling_deformability_with_rigid_blocks}
\end{figure}

\noindent{W}e consider first the geometrically perfect case where the blocks' geometry is exact so there are no initial gaps between the blocks.
Fig. \ref{fig:modeling_deformability_with_rigid_blocks}(b) depicts a two-block assembly of total length $L$, the left block fixed and the right one under in-plane compressive traction $\sigma$. 
The true total shortening depicted in Fig. \ref{fig:modeling_deformability_with_rigid_blocks}(c), which defines the in-plane deformability of the assembly, is the sum of elastic shortenings of the blocks $\Delta_{true} = \frac{\sigma \cdot L}{E}$

In models where contact between deformable blocks in a penalty sense (a common practice in FEM), the total shortening depicted in Fig. \ref{fig:modeling_deformability_with_rigid_blocks}(d) is the sum of the elastic deformations and the interface penetration thus $\Delta_{FEM} = \frac{\sigma \cdot L}{E} + \frac{\sigma}{k_n}$.
Here, the normal penalty parameter $k_n$ is a purely numerical parameter that is usually taken high enough so that the difference between $\Delta_{true}$ and $\Delta_{FEM}$ is negligibly small. 
In our LS-DEM model, which enforces contact between \textit{rigid} blocks in a penalty sense, the total shortening depicted in Fig. \ref{fig:modeling_deformability_with_rigid_blocks}(e) is the interface penetration $\Delta_{LSDEM}=\frac{\sigma}{k_n}$.

By equating $\Delta_{LSDEM}$ to $\Delta_{true}$, it is possible to solve for a deformation-equivalent penalty parameter $k_n=k_n^{def}=\frac{E}{L}$, which ensures that our model reproduces the true total shortening due to the elastic deformation of the blocks, and hence has the same effective in-plane deformablity. This modeling concept, which allows representing deformability under the rigid-body assumption, is the most fundamental one underlying our LS-DEM model for TIS.

The deformation equivalent $k_n^{def}$ can be similarly determined for assemblies with an arbitrary number of blocks.
Consider the assembly of total length $L$ with M+1 blocks shown in Fig. \ref{fig:modeling_deformability_with_rigid_blocks}(f).
The true total shortening is again $\Delta_{true}^{per} = \frac{\sigma \cdot L}{E}$ with the superscript $^{per}$ denoting the geometrically perfect case.
However, the LS-DEM total deformation depicted in Fig. \ref{fig:modeling_deformability_with_rigid_blocks}(h) is now the \textit{sum} of penetrations thus $\Delta_{LSDEM}=\sum \frac{\sigma}{k_n}=M \cdot \frac{\sigma}{k_n}$, where $M$ is the number of interfaces across which penetrations occur.
Equating $\Delta_{true}^{per}$ and $\Delta_{LSDEM}$ and solving for $k_n$ yields the following general expression that applies in the geometrically perfect case:
\begin{equation} \label{eq:k_n_per}
    k_n=k_n^{per} = k_n^{def} = \frac{M \cdot E}{L}
\end{equation}
Unlike the penalty parameters in common FEM/DEM/LS-DEM applications, $k_n^{def}$ is an explicit correlate of $E$, and as such it does not require calibration.
Note that $k_n$ is assumed to be a structural property involving the total number of blocks, the total length, and assuming a common $E$ to all the blocks.  
Cases involving blocks made of different materials or blocks with significantly different dimensions may require reconsidering $k_n$ as block-wise property, and are therefore beyond the present scope.

\noindent{\underline{\textbf{$k_n$ in the geometrically imperfect case}}}

\noindent{I}n real TIS, the geometry of the blocks is never perfect and therefore initial gaps always exist, as depicted in Fig. \ref{fig:modeling_deformability_with_rigid_blocks}(i). 
These gaps increase the effective in-plane deformability, because under in-plane compression, these gaps close without exerting tractions.
In other words, the presence of gaps means a reduced in-plane stiffness and commensurately reduced global stiffness and carrying capacity, see \cite{mirkhalaf_toughness_2019,barthelat_novel_2011}.

Fig. \ref{fig:modeling_deformability_with_rigid_blocks}(j) shows that in the geometrically imperfect case the block deformations and closure of gaps add in-series to true total deformation, which is therefore always larger than in the geometrically perfect case.
Accordingly, we think of the penetration stiffness in the imperfect case $k_n^{imp}$ as the resultant spring of two springs in series - $k_n^{def}$ from Eq. (\ref{eq:k_n_per}), which represents the blocks' deformability, and $k_n^{gaps}$, which represents the contribution of the gaps to the in-plane deformability:
\begin{equation} \label{eq:k_n_imp}
k_n=k_n^{imp}=\frac{k_n^{def} \cdot k_n^{gaps}}{k_n^{def}+k_n^{gaps}}          =\frac{k_n^{def}}{k_n^{def}/k_n^{gaps}+1}
\end{equation}

It can be seen from Eq. (\ref{eq:k_n_imp}) that $k_n^{imp}$ is always smaller than $k_n^{per}$, so the in-plane deformability is always larger in the geometrically imperfect case, as it should be.

In general, the magnitude and distribution of initial gaps is not known a-priori and so there is no close form expression for $k_n^{gaps}$ as the one for $k_n^{def}$.
Therefore, when the effect of imperfections is taken into account using Eq. (\ref{eq:k_n_imp}) - the $k_n^{gaps}$ component of $k_n^{imp}$ requires calibration.

\subsection{Limitations} \label{subsec:Limitations}

\noindent{T}he three main limitations of our model are that (a) it is only applicable to blocks made of relatively rigid materials for which $k_n$ is sufficiently large and the penetrations sufficiently small; soft materials with too small $k_n$'s may induce too large penetrations that could overly distort the actual (penetration-less) kinematics; (b) it does not account for material non-linearity, specifically fracture, which sometimes plays a role in TIS' failure; and (c) it does not provide a resolution of the bulk stresses, which are usually of interest in design\footnote{Nevertheless, bulk stresses can be estimated at post-processing by solving the continuum problem of blocks loaded by the contact surface tractions which our model provides. This can be done using any continuum model, e.g., FEM.}.

\section{Results and discussion} \label{sec:Results and discussion}
The structural configurations considered in this manuscript and their LS-DEM models are described in section \ref{subsec:Configuration}. 
The model is validated through comparisons with experimental results and FEM analyses from \cite{mirkhalaf_toughness_2019} in section \ref{subsec:Validation}.

\subsection{Configuration} \label{subsec:Configuration}
All the numerical examples in this manuscript consider centrally-loaded square TIS slabs studied in \cite{mirkhalaf_toughness_2019}.
This configuration was chosen for three reasons. 
First, centrally loaded slabs are the most common TIS studied in the literature. 
Second, \cite{mirkhalaf_toughness_2019} contains detailed experimental information and ample data for comparison and validation. 
Third, the polyhedral blocks used in \cite{mirkhalaf_toughness_2019} have planar faces and therefore they interact across matching planes. 
Such interfaces represent the simplest form of conforming contacts, as distinct from the non-conforming contact typically modeled with discrete element methods. 
As such, they are a natural starting point for a future investigation of more complex cases of conforming contacts that characterize TIS with curved-face (e.g., osteomorphic) blocks \cite{dyskin_fracture_2003,djumas_deformation_2017,djumas_enhanced_2016,estrin_design_2021}.

\noindent{\underline{\textbf{Experimental set-up}}}

\begin{figure} 
    \includegraphics[width=1\textwidth]{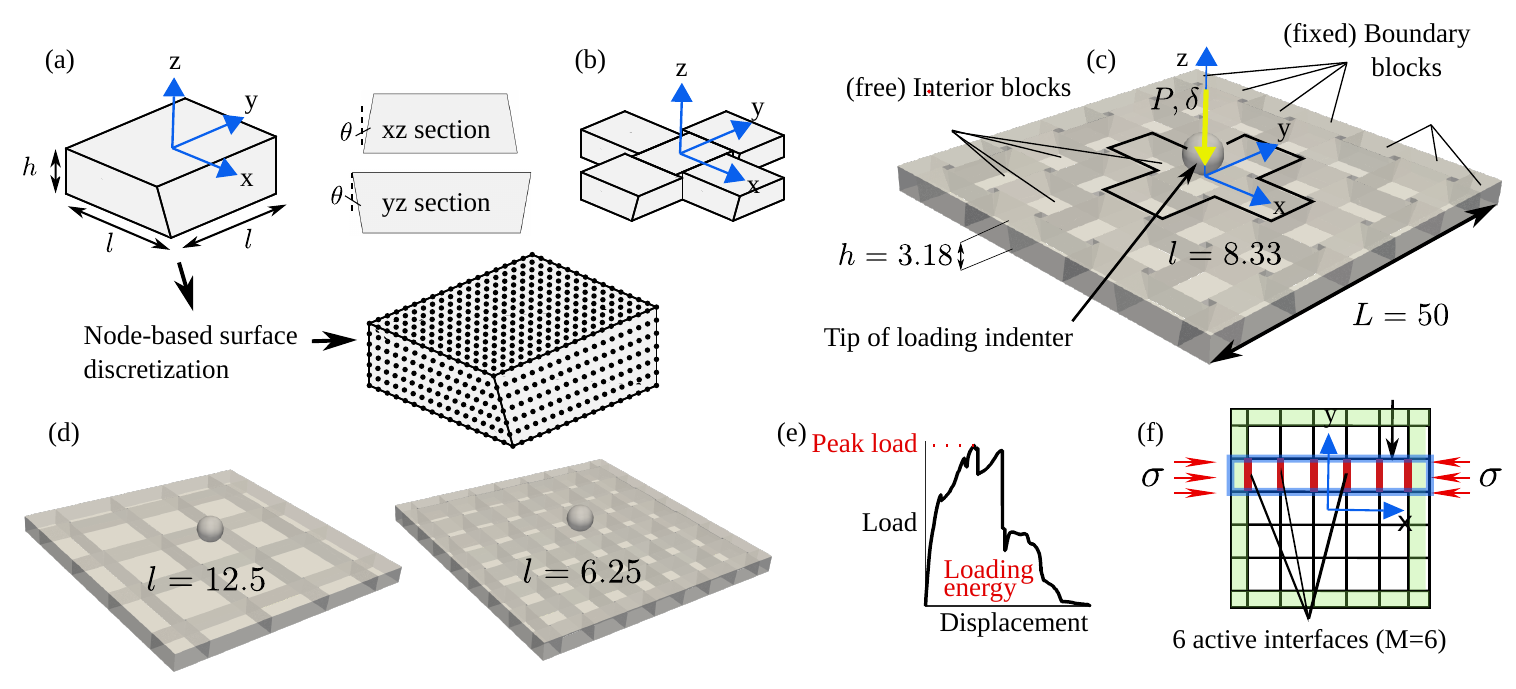} 
    \centering
    \caption{Configuration - (a) a typical internal block, its two cross-sections, and its LS-DEM surface discretization; (b) a basic five-block interlocked cell; (c) The full 5x5 slab and its boundary conditions; (d) the 3x3 and 7x7 slabs; (e) a typical load displacement curve; and (f) a typical strip to determining $k_n^{def}$ from Eq. (\ref{eq:k_n_per})}
    \label{fig:configuration}
\end{figure}

\noindent{Fig. \ref{fig:configuration}(a)} shows the truncated polyhedral block used in \cite{mirkhalaf_toughness_2019} and its xz and yz cross-sections.
The bottom face of the blocks is a square with side length $l$, and the angle of inclination of its sloping lateral faces is $\theta$.
Fig. \ref{fig:configuration}(b) shows the basic 5-block cell formed by surrounding a block by four similar ones rotated with respect to it by 90$^{\circ}$ about the $z$ axis.
Fig. \ref{fig:configuration}(c) shows an entire slab with the contour of a basic cell around the central block marked in black.
The slabs' dimensions are 50 x 50 x 3.18 mm, and they consist of boundary blocks along the edges and internal blocks.
The boundary blocks are either halves or quarters (in the four corners) of the internal blocks in a way that the assembled slab's convex hull is a straight parallelepiped.

Slabs with identical overall dimensions but with three block sizes - medium, large, and small - are considered.
The medium-block slab, depicted in Fig. \ref{fig:configuration}(c), has 5x5=25 internal blocks with l=8.33 mm, and it is referred to as the 5x5 slab.
The large-, and small-block slabs, depicted in Fig. \ref{fig:configuration}(d), have, respectively, 3x3=9 and 7x7=49 internal blocks with l=12.5 and l=6.25 mm, and they are referred to as the 3x3 and 7x7 slabs.

The slabs in \cite{mirkhalaf_toughness_2019} were confined by a stiff peripheral frame that held the boundary blocks in place without pre-compression.
They were quasi-statically loaded by a pin indenter that pushed the central block in the negative z direction at a rate of 0.01 mm/sec.
The force $P$ exerted by the indenter on the slab and the corresponding indenter displacement $\delta$ are indicated by a yellow arrow in the -z direction in Fig. \ref{fig:configuration}(c).
Fig. \ref{fig:configuration}(e) shows an experimental $P-\delta$ curve, with the main global response parameters indicated in red.

\noindent{\underline{\textbf{LS-DEM model}}}

\noindent{The} node-based surface discretizations in the LS-DEM model is illustrated in Fig. \ref{fig:configuration}(a) for the central block with $\theta=10^{\circ}$\footnote{More details on the level-set geometrical representation of the blocks, which are used in LS-DEM to evaluate the contact penetrations, can be found in \cite{kawamoto_level_2016}}.

First, the blocks were positioned and oriented in the initial undeformed configuration of the slab as illustrated in Fig. \ref{fig:configuration}(c), and the boundary conditions were affected by fixing the boundary blocks. 
Next, the assembly was subjected to gravity until it reached a relaxed state, i.e., until the kinetic energy lowered to effectively zero.
The relaxed positions and rotations of the blocks were then taken as the initial conditions for the main loading phase - the indentation.

For the indentation loading, the 2.5 mm spherical tip of the indenter was modeled, see Fig. \ref{fig:configuration}(c,d), and a constant velocity in the negative z direction was prescribed to it.
To expedite the analyses, the loading speed was taken as high as possible, but always low enough to avoid inertial effects.
The loading rate values ranged between 3-6 mm/sec.
The density of the alumina-silicate blocks was taken equal to $2.5\cdot10^{-6}\frac{kg}{mm^3}$, and a friction coefficient $\mu$=0.23 was used, in accordance with \cite{mirkhalaf_toughness_2019}.

Numerical tests detailed in \cite{feldfogel_discretization-convergent_2022} were made to determine the refinement of the surface discretization and of the level-set geometrical representation of the blocks necessary for numerical convergence of the results.
The converged surface discretization and Level-set parameters was found to be 0.6 mm and 0.25 mm, respectively.
These values were used for all the analyses in this manuscript.
Additional information regarding the run time and computational cost of the analysis is provided in Appendix \ref{appendix}.

\subsection{Validation} \label{subsec:Validation}

The model's validity is assessed in two parts through comparisons with the experimental and numerical results  reported in \cite{mirkhalaf_toughness_2019}.
First, the non-calibrated $k_n^{per}$ from Eq. \ref{eq:k_n_per} is used, under the assumption of a geometrically perfect case. 
This idealized assumption was also made in the FEM analyses in \cite{mirkhalaf_toughness_2019}, which allows to compare LS-DEM to FEM. 
Second, the more realistic geometrically imperfect case is assumed, which requires calibration of the $k_n^{gaps}$ component of $k_n^{imp}$, as discussed in relation to Eq. (\ref{eq:k_n_imp}). This part allows to directly assess and examine the ability of the model to capture and match experimental observations. 

\noindent{\underline{\textbf{Non-calibrated model}}}

\noindent{L}SDEM's ability to estimate the response of the centrally loaded slabs in \cite{mirkhalaf_toughness_2019} is compared with the FEM analyses therein, under the assumption of a geometrically perfect case, common to both models. 
In accordance with the assumption of a geometrically perfect case, we used $k_n^{per}$ values of 1.5, 2.25, and 3 GPa/mm for the 3x3, 5x5, and 7x7 slabs, respectively.
\begin{comment}
These values were obtained by substituting $E=18.7$ GPa from \cite{mirkhalaf_toughness_2019}, $L=50$ mm, and $M=4,6,8$ for the 3x3, 5x4, and 7x7 slabs, respectively, in Eq. (\ref{eq:k_n_per}).
\end{comment}

In \cite{mirkhalaf_toughness_2019}, only the $\theta$=2.5$^{\circ}$ were analyzed.
For all three slabs, the FEM analyses overestimated the experimental peak load and loading energy, see Table \ref{Table:1}.
\begin{comment}
The initial stiffness, which was also over estimated by a factor of 1.5-3 is not included in Table \ref{Table:1} because it is not explained in \cite{mirkhalaf_toughness_2019} how exactly it was calculated and therefore the basis for comparison with our model is missing.
\end{comment}
Similarly to the FEM analyses in \cite{mirkhalaf_toughness_2019}, our model also over-estimates the response parameters, see Table \ref{Table:1}. 

Comparing columns 3 and 4 in Table \ref{Table:1}, the largest LS-DEM overestimation factor for each response parameter is much smaller than the smallest corresponding FEM factor.
This means that, even without calibration, LS-DEM provides much closer estimate of the failure response parameters than FEM, supporting its validity.

\begin{table}
\begin{center}
\begin{tabular}{|c|c|c|c|} 
  \hline
  \multirow{2}{*}{Response parameter} & \multirow{2}{*}{Assembly index} & \multicolumn{2}{|c|}{Overestimation factor}\\
  \cline{3-4}
  & & FEM\cite{mirkhalaf_toughness_2019} & LS-DEM with $k_n^{per}$\\
  \hline\hline
  \multirow{3}{*}{peak load [N]} & 3x3 & \multirow{3}{*}{14-15} & 7.1\\
  & 5x5 & & 3.7\\
  & 7x7 & & 3.4\\
  \hline
  \multirow{3}{*}{Loading energy [N$\cdot$mm]} & 3x3 & \multirow{3}{*}{9-13} & 6.2\\
  & 5x5 & & 5.6\\
  & 7x7 & & 6.0\\
  \hline
\end{tabular}
\captionsetup{justification=centering}
\caption{Factors of error of adapted LS-DEM and FEM relative to \cite{mirkhalaf_toughness_2019} experiments in the geometrically perfect case - our model is more accurate than FEM, even without parameter calibration}
\label{Table:1}
\end{center}
\end{table}

As did Mirkhalaf et al. \cite{mirkhalaf_toughness_2019}, we attribute the over-estimation of the structural response parameters to initial gaps that reduced the experimental slabs' stiffness and strength, but which are not accounted for in analyses using $k_n^{per}$.
\begin{comment}
\footnote{From \cite{mirkhalaf_toughness_2019}: "We attributed these differences (between FEM and the experiments) to presence of small gaps between the blocks resulted from statistical variations in the shape of blocks, an effect which has been previously found to significantly affect the mechanical performance in similar materials (Barthelat and Zhu, 2011)"}. 
\end{comment}
Our model can account for such gaps through the $k_n^{gaps}$ term in $k_n^{imp}$, under the geometrically imperfect case, see Eq. (\ref{eq:k_n_imp}).
The ability of the model to obtain a closer agreement with-, and to predict the  experimental results in \cite{mirkhalaf_toughness_2019} using calibrated $k_n^{gaps}$ is discussed next.

\noindent{\underline{\textbf{Calibrated model}}}

\noindent{T}o calibrate the model, we first explore the effects of geometrical imperfections on the structural response through $k_n=k_n^{imp}$ that are smaller than $k_n^{per}$. 
The $\theta=5^\circ$ 5x5 slab was chosen as the calibration slab because it is the intermediate one in terms of number of blocks and $\theta$.

Next, based on optimal agreement of the model with the experimental benchmark in terms of the $P-\delta$ curve and the failure mechanism, we choose the best fit $k_n$, and solve for the calibrated $k_n^{gaps}$ from Eq. (\ref{eq:k_n_imp}), using the known $k_n^{def}$=2.25 GPa/mm for the 5x5 slab.

Last, using the calibrated $k_n^{gaps}$, we validate the model with eight validation slabs - the remaining combinations of the 3x3, 5x5, and 7x7 slabs with $\theta=2.5^\circ, 5^\circ, 7.5^\circ$\footnote{Slabs with $\theta=10^{\circ}$ were not included among the eight validation slabs because their failure was reported in \cite{mirkhalaf_toughness_2019} to have been affected by material damage, which our model does not account for.}
\begin{comment}
- "In these cases ($\theta=10^{\circ}$), the interlocking stresses are too high, which leads to excessive block damage"
\end{comment}

\begin{figure}[H] 
    \includegraphics[width=1\textwidth]{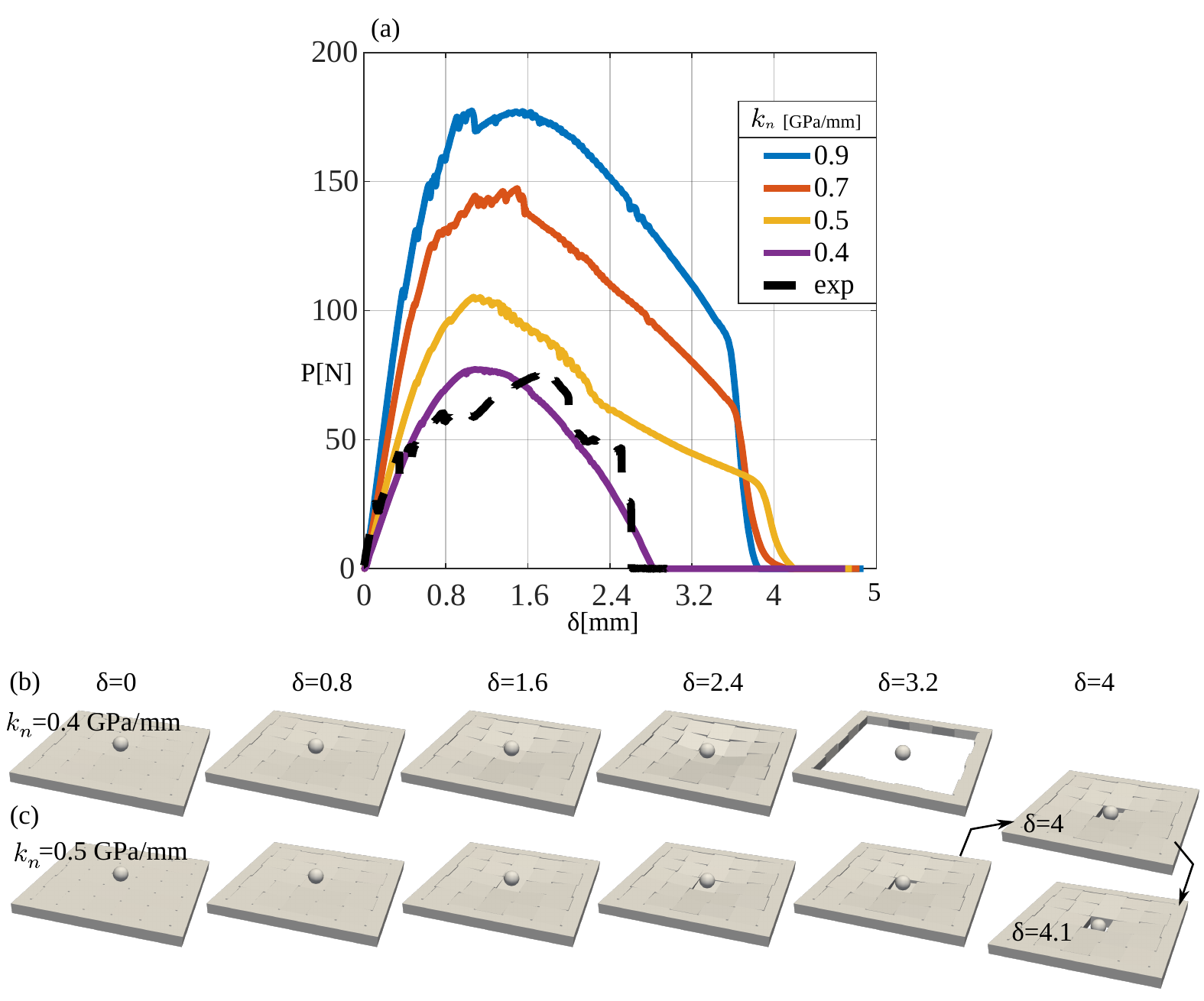} 
    \centering
    \caption{Calibrated model: (a) $P-\delta$ curves of the $\theta=5^{\circ}$ 5x5 slab with $k_n$ smaller than $k_n^{per}$; (b-d) failure mechanism snapshots for selected $k_n$'s.}
    \label{fig:knCalibration}
\end{figure}

Fig. \ref{fig:knCalibration}(a) depicts $P-\delta$ curves for the $\theta=5^{\circ}$ 5x5 slab with $k_n$'s ranging from $k_n^{per}$=0.9 GPa/mm down to 0.4 GPa/mm.
The $k_n$=0.4 GPa/mm and $k_n$=0.5 GPa/mm curves are seen to envelope the experimental one.
The $k_n$=0.4 GPa/mm curve is in very close agreement in terms of peak load, loading energy, and ultimate displacement.
With $k_n$=0.5 GPa/mm, these parameters are not as closely estimated, but the model more accurately captures the slab's linear response and initial stiffness.

Fig. \ref{fig:knCalibration}(b,c) depicts the failure snapshots for the $k_n$=0.4 GPa/mm and $k_n$=0.5 GPa/mm.
For $k_n$=0.5 GPa/mm, the mechanism follows closely the three response stages reported in \cite{mirkhalaf_toughness_2019}: (1) The slab starts bending as a whole with the blocks initially sticking; (2) the loaded central blocks starts slipping; (3) the slipping of the loaded block become more dominant until, in the end, it falls off while the rest of the assembly partially recovers the deformation by rebounding upwards\footnote{"While the center block is being pushed out and the deformation localizes, the force decreases, and the rest of the panel partially recovers its deformation \cite{mirkhalaf_toughness_2019}".}.
In contrast, in the $k_n$=0.4 GPa/mm case the blocks stick throughout and the slab collapses globally with all the blocks eventually falling off.

Based on the model's ability to capture with $k_n$=0.5 GPa/mm all the experimental response stages reported in \cite{mirkhalaf_toughness_2019}, while being reasonably close to the peak load, it was selected over the $k_n$=0.4 GPa/mm case as the best fit $k_n$. Substituting $k_n$=0.5 GPa/mm and $k_n^{def}$=2.25 GPa/mm in Eq. (\ref{eq:k_n_imp}) and solving for $k_n^{gaps}$ yields the calibrated $k_n^{gaps}$=0.65 GPa/mm.

\begin{comment}
\begin{table}[h!]
\begin{center}
\begin{tabular}{ | c | c || c | c || c | c |} 
  \hline
  \multirow{2}{*}{Assembly} & \multirow{2}{*}{$\theta$ [$^\circ$]} &  \multicolumn{2}{|c||}{$P_{max}^{exp}/P_{max}^{model}$} & \multicolumn{2}{|c|}{$LE_{max}^{exp}/LE_{max}^{model}$}\\
  \cline{3-6}
  & & $k_n^{*}=5e5$ & $k_n^{*}=4.5e5$ & $k_n^{*}=5e5$ & $k_n^{*}=4.5e5$\\
  \hline\hline
  7x7 & \multirow{3}{*}{2.5} & 0.85 & 0.70 & 1.02 & 0.79\\
  5x5 &                      & 1.36 & 1.16 & 1.42 & 1.72\\
  3x3 &                      & 1.66 & 1.52 & 1.44 & 1.15\\
  \hline
  7x7 & \multirow{3}{*}{5}   & 0.88 & 0.72 & 1.25 & 0.68\\
  5x5 &                      & \textbf{1.40} & \textbf{1.24} & \textbf{1.84} & \textbf{1.26}\\
  3x3 &                      & 1.74 & 1.23 & 2.13 & 1.69\\
  \hline
  7x7 & \multirow{3}{*}{7.5} & 1.24 & 0.97 & 1.92 & 0.99\\
  5x5 &                      & 1.21 & 1.54 & 1.37 & 1.71\\
  3x3 &                      & 1.58 & 1.36 & 2.22 & 1.88\\
  \hline\hline
  \multicolumn{2}{|c||}{Average error factors} & \textbf{1.32} & \textbf{1.16} & \textbf{1.62} & \textbf{1.32}\\
  \hline\hline
  \multicolumn{2}{|c||}{Standard deviation} & 0.32 & 0.31 & 0.42 & 0.45\\
  \hline\hline
  \multicolumn{2}{|c||}{Standard deviation [\%]} & 24 & 27 & 26 & 34\\
  \hline\hline
\end{tabular}
\captionsetup{justification=centering}
\caption{The adapted LS-DEM is much more accurate than FEM \cite{mirkhalaf_toughness_2019}}
\label{Table:calibrated model error factors}
\end{center}
\end{table}
\end{comment}

\begin{figure}[H] 
    \includegraphics[width=1\textwidth]{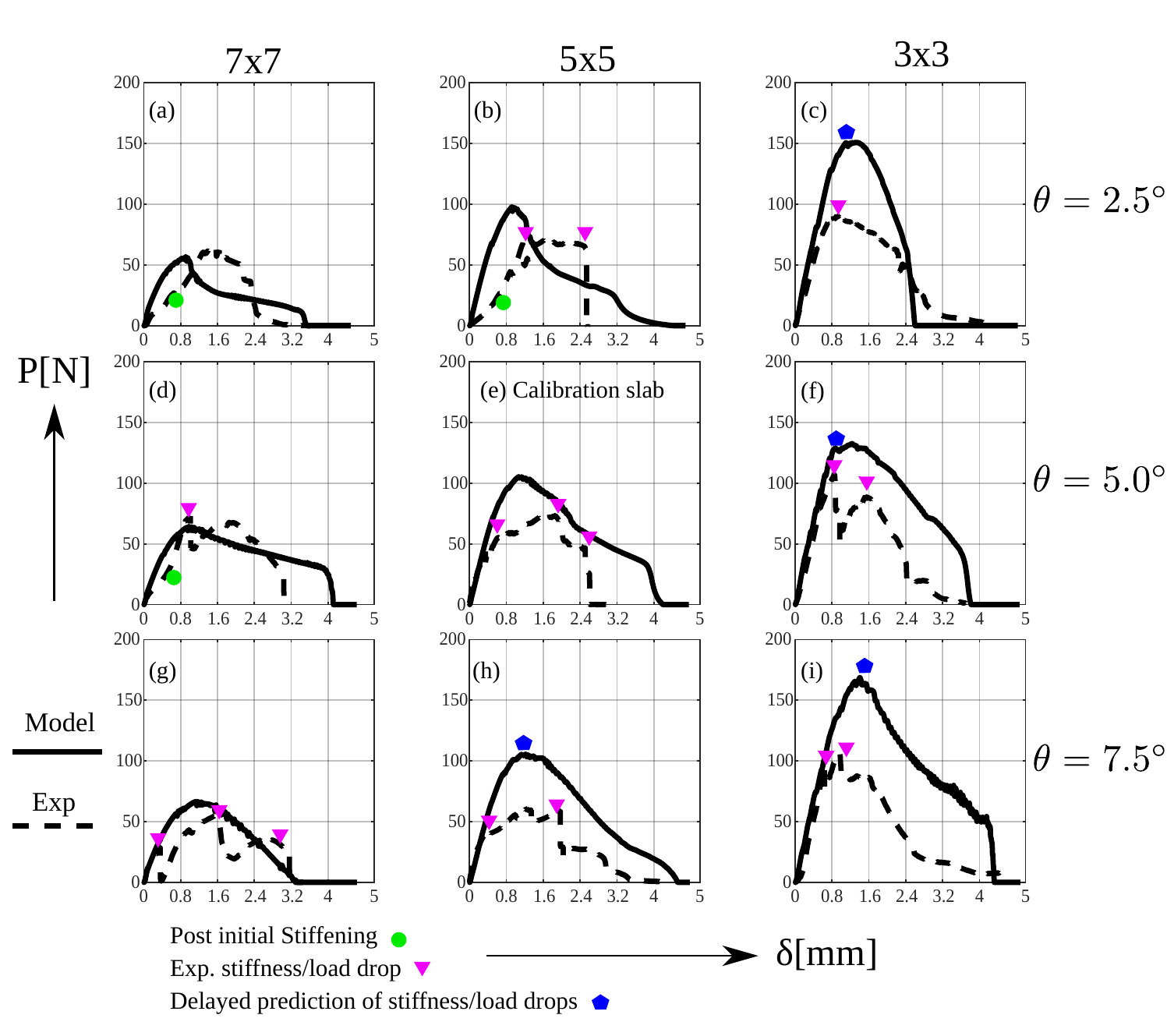} 
    \centering
    \caption{The calibrated model's $P-\delta$ curves are close to  \cite{mirkhalaf_toughness_2019} across the eight validation slabs (a-d) and (f-i)}
    \label{fig:validation_50_Pdelta}
\end{figure}

\noindent{\underline{\textbf{Validation}}}

\noindent{With} the calibrated $k_n^{gaps}$=0.65 GPa/mm, we move on to validate the LS-DEM model by comparing the response of the eight validation slabs to the experimental results. 
For the 3x3 and 7x7 slabs, $k_n$ is 0.45 GPa/mm and 0.53 GPa/mm, respectively, from Eq.(\ref{eq:k_n_imp}).

Fig. \ref{fig:validation_50_Pdelta} depicts the $P-\delta$ curves of the nine slabs, obtained with the calibrated $k_n^{gaps}$=0.65 GPa/mm.
It shows that the calibrated model: (1) is close to the experimental benchmark across the validation slabs; (2) correctly captures the increase in peak load and loading energy with block size for all $\theta$'s; (3) quantitatively matches the initial stiffness across the different block sizes and $\theta$'s (to a lesser degree in cases a,b,d); and (4) captures the negative stiffness phases in (d,f,h,i) and the load drops (c,d,e,f,i) at final stages of failure.

\begin{figure}[H] 
    \includegraphics[width=1\textwidth]{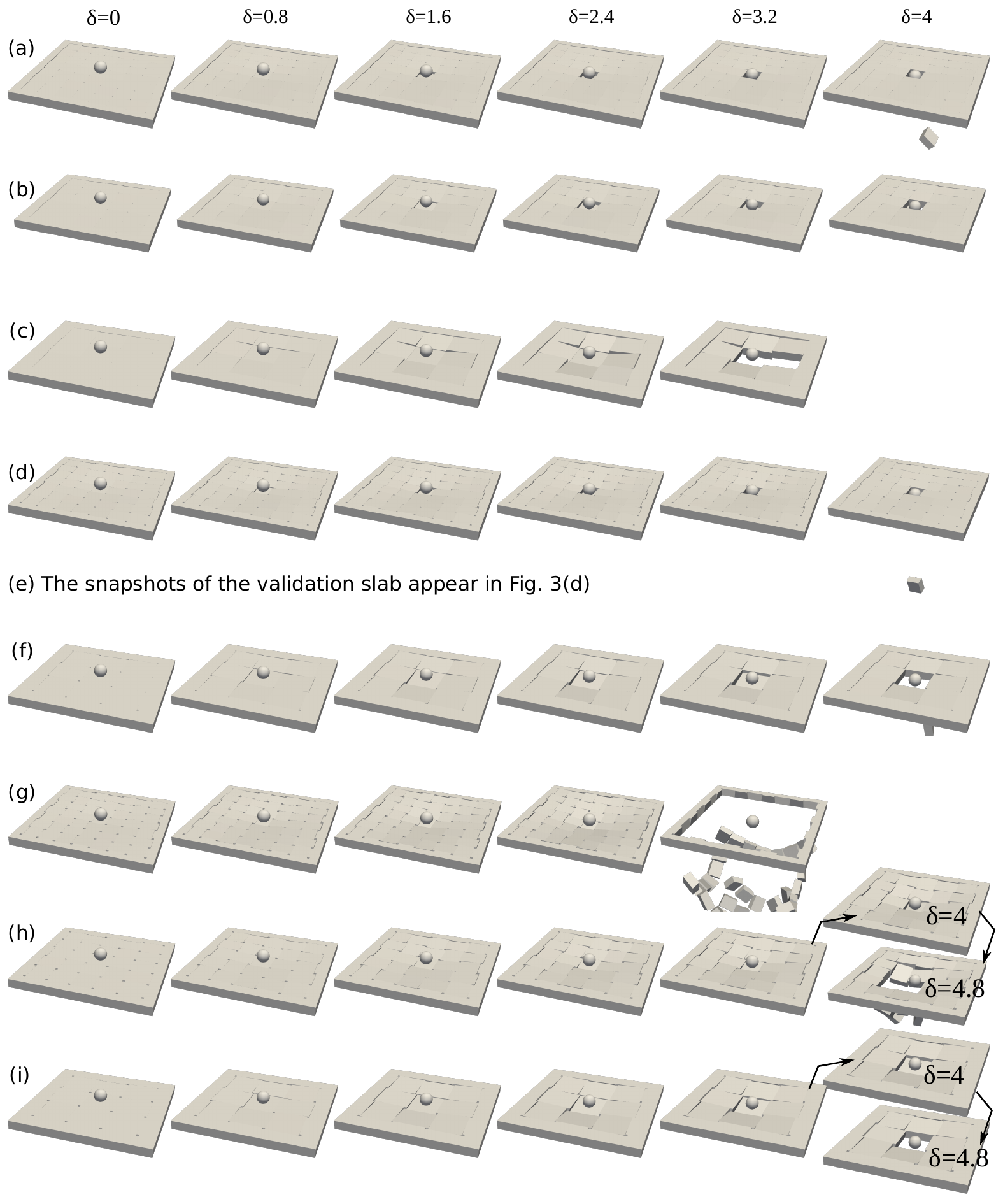} 
    \centering
    \caption{Failure mechanism snapshots of the eight validation slabs}
    \label{fig:validation_50_snapshots}
\end{figure}

Fig. \ref{fig:validation_50_snapshots} depicts the failure mechanism for the eight validation cases corresponding to the $P-\delta$ curves in Fig. \ref{fig:validation_50_Pdelta}(a-d,f-i).
It shows that in most cases the calibrated model correctly captures the experimentally observed slip-governed failure mechanism, with the central block slipping out of the assembly and with some upward rebound \cite{mirkhalaf_toughness_2019}.
Lesser agreement is obtained in cases (c) and (h), where more than one block eventually fall off, and a stick-governed failure different from the experimentally reported one was obtained in case (g).

Table \ref{Table:calibrated model error factors} summarizes the error factors defined as the ratios between the experimental and the computed peak loads $P_{max}^{exp}/P_{max}^{model}$ and loading energies $LE_{max}^{exp}/LE_{max}^{model}$ for the nine slabs.
It shows that: (1) the error factors are smaller by an order of magnitude compared with the FEM analyses in \cite{mirkhalaf_toughness_2019}; (2) while the calibrated model somewhat overestimates the loading energy in some of the slabs, it provides close estimates of the peak load, which is usually the more important parameter, across the validation cases; (3)
the largest error factor of 2.22 is still a workable estimate, given common safety factors for structures; and, importantly (4) the error factors of the calibration case are similar to the average error factors of the validation cases.

\begin{table}[H]
\begin{center}
\begin{tabular}{ | c | c || c || c |} 
  \hline
  Assembly & $\theta^\circ$ & $P_{max}^{exp}/P_{max}^{model}$ & $LE_{max}^{exp}/LE_{max}^{model}$\\
  \hline\hline
  7x7 & \multirow{3}{*}{2.5} & 0.85 & 1.02\\
  5x5 &                      & 1.36 & 1.42\\
  3x3 &                      & 1.66 & 1.44\\
  \hline
  7x7 & \multirow{3}{*}{5}   & 0.88 & 1.25\\
  5x5 &                      & \textbf{1.40} & \textbf{1.84}\\
  3x3 &                      & 1.74 & 2.13\\
  \hline
  7x7 & \multirow{3}{*}{7.5} & 1.24 & 1.92\\
  5x5 &                      & 1.21 & 1.37\\
  3x3 &                      & 1.58 & 2.22\\
  \hline\hline
  \multicolumn{2}{|c||}{Average error factors} & \textbf{1.32} & \textbf{1.62} \\
  \hline\hline
  \multicolumn{2}{|c||}{Standard deviation} & 0.32 & 0.42\\
  \hline\hline
  \multicolumn{2}{|c||}{Standard deviation [\%]} & 24 & 26\\
  \hline\hline
\end{tabular}
\captionsetup{justification=centering}
\caption{Error factors in validation of calibrated model - the errors are smaller than FEM's by an order of magnitude, reasonable in absolute terms, and similar in the calibration slab and the validation slabs, supporting the model's validity}
\label{Table:calibrated model error factors}
\end{center}
\end{table}

The ability of the calibrated model to correctly capture the failure mechanism, closely approximate the experimental $P-\delta$ curves, estimate the response parameter to a good engineering precision across a wide range of experimental slabs, and correctly capture experimentally observed parametric trends support its validity.
Most importantly, it proves that modeling deformability under the rigid-body assumption as discussed in section \ref{sec:Methodology} allows capturing the essentials of the behavior and failure of TIS.
This strongly supports the basic concept underlying our modelling approach.

\begin{comment}
\noindent{\underline{\textbf{Caveat}}}

\noindent{N}onwithstanding the calibrated model's ability to provide good results, we recognize that $k_n^{imp}$ combines independent (yet related) physical phenomena, namely block deformability and initial gaps, that reduce the in-plane stiffness.
While this allows to obtain a much improved agreement with the experimental benchmark, it does not allow to quantify the effects of each of these features (initial gaps and boundary conditions imperfections) separately. 
For such a separate quantification, additional (and currently unavailable) experimental information and an elaborate numerical study that are beyond the scope of this manuscript would be required.

Along the same lines but from a broader perspective, it makes methodological sense to disentangle all the ingredients that go into $k_n^{imp}$, including the block deformability, so that the latter reverts to being a purely numerical parameter penalizing penetrations and loses its current dual nature. 
Such an approach would allow a cleaner and more detailed investigation of the separate effects of each the different ingredients that currently make up $k_n^{*}$.
It would entail directly accounting for block deformability independently of $k_n^{*}$, which is reserved for future work.

\end{comment}

\noindent{\underline{\textbf{Discrepancies}}}

\noindent{W}e attribute some of the discrepancies between the experimental results and the model predictions to experimental imperfections not taken into account by the model.
These may have included a single dominant initial gap, compliance of the peripheral boundary element that kept the boundary blocks in place, or small in-plane slipping failures between the peripheral element and the boundary blocks.
Such factors may explain the markedly smaller initial stiffness and unexplained stiffening in cases (a,b,d), indicated by green circles in \ref{fig:validation_50_Pdelta}.
That such imperfections indeed caused the reduced initial stiffness is supported the fact that: (1) good agreement with the experiments was obtained in cases (c),(e),(f),(g, after the initial early load drop), (h) and (i); and (2) in all these cases the variations in the initial stiffness were relatively small.

Aside from experimental-imperfections-induced discrepancies, Fig. (\ref{fig:validation_50_Pdelta}) reveals a repeated discrepancy which we attribute to our model's overly simplistic friction modeling.
It is that, in our model, the load/stiffness drops that follow the initial linear response (and which are indicated by blue five-point stars in the figure) occur later than in the experimental curves.
This feature explains the general overestimation of the response by the calibrated model, and specifically for the 3x3 slabs where it is most pronounced, and translates to larger error factors due the sharper experimental drops.

The delayed stiffness/load drop transition of the model is attributed to three elements, the last two of which are the more relevant ones:
(1) high interlocking stresses and material damage associated with larger blocks and higher $\theta$'s were reported in \cite{mirkhalaf_toughness_2019} to precipitate the push out of the central block and thereby to reduce the slabs' carrying capacity.
This specifically explains the poorer agreement for the 3x3 and the ($\theta=7.5^{\circ}$) slabs.
(2) The inherently unstable drop in friction force at the stick-slip transition and its translation to load drops and instabilities at the global level are not accounted for in our simple bi-linear Coulomb friction law. The sharp, and otherwise unexplained load/stiffness drops indicated by pink triangles throughout the $P-\delta$ curves in Fig. \ref{fig:validation_50_Pdelta}(d-i) suggest that friction-slip induced instabilities indeed were at play; and (3) the inevitable variability of the friction coefficient in the experiments was not hitherto taken into account in the analyses, where a fixed $\mu=0.23$ was used.  

\begin{figure}[H] 
    \includegraphics[width=0.8\textwidth]{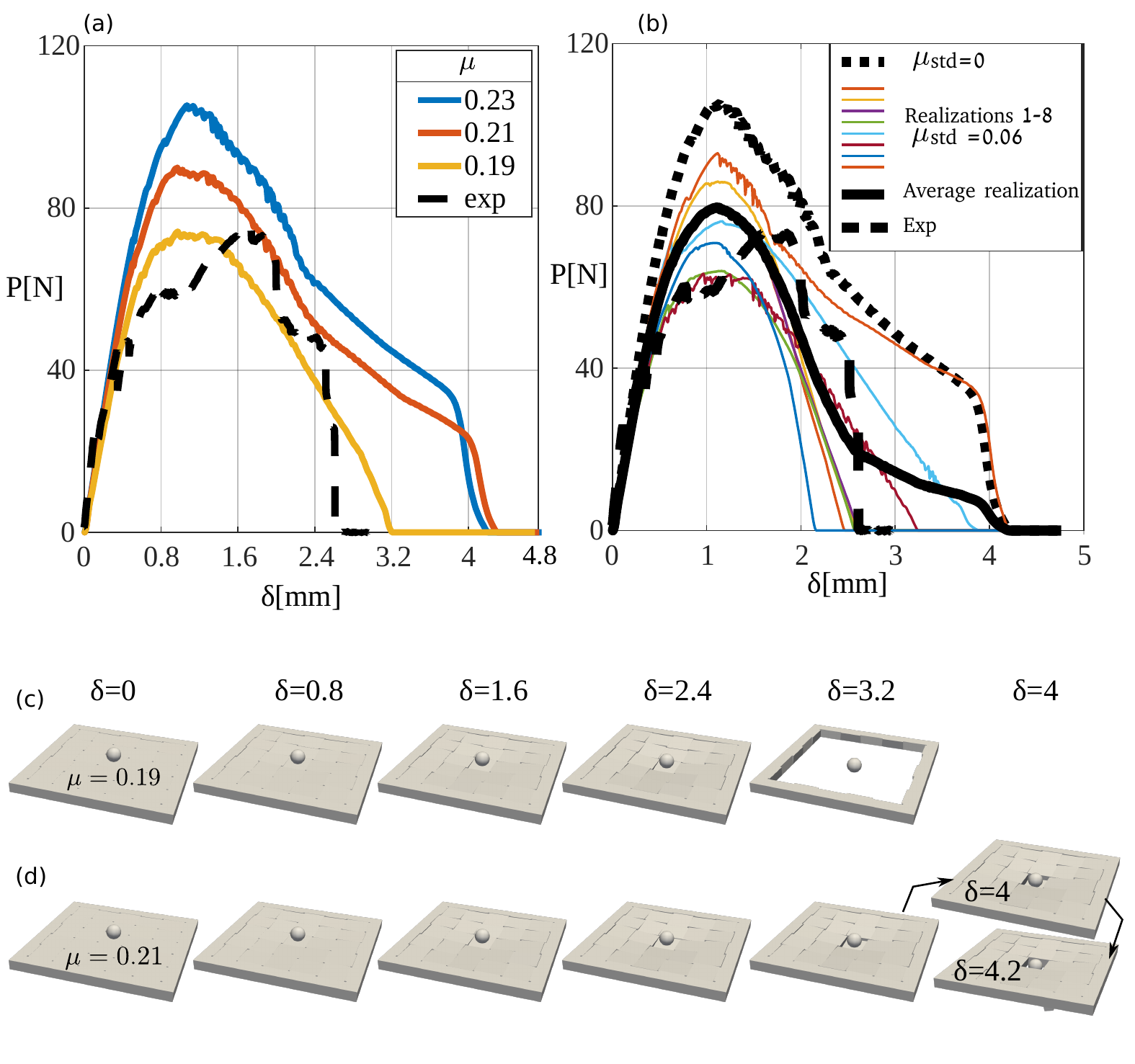} 
    \centering
    \caption{The discrepancies between analyses and experiment can be reduced by simple modifications in the friction model: (a) $P-\delta$ curves with smaller-than-nominal $\mu$'s; (b) $P-\delta$ curves from 8 realizations with randomly generated $\mu$'s; (c,d) failure mechanism snapshots for selected $\mu$'s.}
    \label{fig:discrepancies}
\end{figure}

While fully addressing the role of experimental imperfections, material damage or fracture, and explicitly accounting for the friction-slip associated force drop is beyond the scope of this manuscript, some frictional strength aspects can be approximated in our model. 
Specifically, the friction force drop upon sliding initiation can be roughly  approximated by assuming smaller friction coefficients than the nominal one. The effects of $\mu$ variability can be examined by attributing a random distribution of friction coefficients to the blocks. Both these possibilities are explored next in Fig. \ref{fig:discrepancies} in the context of the $\theta=5^{\circ}$ 5x5 slab.

Fig. \ref{fig:discrepancies}(a) shows that smaller friction coefficients contribute to an earlier stiffness/load drop leading to a closer agreement with the experimental curve.
However, while the failure mechanism for the $\mu=0.21$ remains correct, see Fig. \ref{fig:discrepancies}(d), the better $P-\delta$ agreement with $\mu=0.19$ comes at the expense of losing the correct failure mechanism, see Fig. \ref{fig:discrepancies}(c).

Fig. \ref{fig:discrepancies}(b) illustrates the effect of introducing $\mu$ variability to the model. 
The eight thin colored lines correspond to eight realizations of the model wherein the $\mu$'s for the different blocks were obtained randomly from a normal distribution with mean 0.23 (the nominal value) and standard deviation (std) 0.06.
It can be seen that the average realization enveloped by the eight realizations is much closer to the experimental benchmark than the reference analysis with deterministic $\mu=0.23$.
Fig. \ref{fig:discrepancies}(a,b) show that with simple approximate modifications to the nominal friction coefficient, still better agreement with the experimental results can be obtained, supporting the reasoning given to the discrepancies.

\section{Conclusion} \label{sec:Conclusion}

In this study, we have presented a new modeling approach for the behavior and failure of topologically interlocked structures.
This approach, based on the Level-Set-Discrete-Element-Method  aims at providing better descriptive and predictive capabilities for the their complex behavior and failure

After outlining the theoretical basis of our approach, we have shown that our model correctly captures and predicts experimentally observed slip-governed failure in centrally loaded topologically interlocked slabs, and that it can estimate the key response parameters much more closely than presently available models.

The theoretical basis we have outlined and the improved ability of our model to describe and predict the behavior of topologically interlocked structures establish the proof-of-concept of our new Level-Set-DEM approach.

\begin{comment}
Having established the soundness of the LS-DEM model in the highly challenging context of TIS, where geometrical irregularity, complex contact interactions, large-displacement and slip-governed failure, topological interlocking, and, critically, block deformability - all play a major role, we believe that our model can be suitable to study many other structural systems applications based on discrete building blocks. 
These range from other TI forms such as shells, through masonry panels, to compression-only structures vaults.
This avenue is reserved for future research.
\end{comment}

\section{Acknowledgement}
Shai Feldfogel was a Swiss Government Excellence Scholarship holder for the academic years 2021-2022 (ESKAS No. 2021.0165).

\appendix 
\section{Appendix - Computational information} \label{appendix}

The c++ LS-DEM code used for the analyses in this manuscript was run on the ETH Euler cluster.
A typical analysis of the 5x5 assembly (49 blocks + the spherical indenter tip) with the most refined surface discretization with a distance of 0.06 mm between surface nodes took approximately 50 CPU hours to run, without parallelization and code optimization.
The preprocessing stage, where the Level-Set geometrical representations of the blocks are calculated, and which only has to be done once per structure, took about 8 hours. The 100,000 time-increments of the relaxation under gravity took about 7 hours, and the 500,000 time-increments of indentation loading took about 35 hours.
The time step $\Delta t_{LSDEM}$ required for numerical stability of the explicit formulation was about 1 ms.
Analyses with less refined discretizations that yielded results fairly close to the converged ones took only a few hours.

\printbibliography

\end{document}